\documentclass[a4paper,onecolumn,11pt,unpublished]{quantumarticle}
\pdfoutput=1
\usepackage[T1]{fontenc}
\usepackage[utf8]{inputenc}
\usepackage[english]{babel}
\usepackage{amsmath,amssymb,amsthm}
\usepackage{etoolbox}
\usepackage{microtype}
\usepackage{booktabs,tabularx}
\usepackage[numbers]{natbib}
\usepackage{xurl}
\usepackage{xcolor}
\usepackage[colorlinks=true,allcolors=quantumviolet,pagebackref=true]{hyperref}
\usepackage{bookmark}
\usepackage{graphicx}

\AtBeginDocument{%
  \renewcommand*{\doi}[1]{doi: \begingroup\urlstyle{rm}%
    \href{https://doi.org/#1}{\nolinkurl{#1}}\endgroup}}
\renewcommand*{\backref}[1]{}
\renewcommand*{\backrefalt}[4]{%
  \ifcase #1\relax
  \or\space Cited on page~#2.%
  \else\space Cited on pages~#2.%
  \fi}

\newcommand{\statementqedsymbol}{\ensuremath{\square}}
\newcommand{\proofqedsymbol}{\ensuremath{\blacksquare}}

\newcommand{\statementqed}{%
  \par\nobreak\noindent\statementqedsymbol\par
}

\makeatletter
\renewenvironment{proof}[1][\proofname]{%
  \par\pushQED{\proofqed}%
  \normalfont\topsep6\p@\@plus6\p@\relax
  \trivlist
  \item[\hskip\labelsep\itshape #1\@addpunct{.}]\ignorespaces
}{%
  \par\nobreak\noindent\popQED
  \endtrivlist\@endpefalse
}
\newcommand{\proofqed}{\proofqedsymbol}
\makeatother

\theoremstyle{plain}
\newtheorem{theorem}{Theorem}[section]

\newtheorem{proposition}[theorem]{Proposition}
\newtheorem{corollary}[theorem]{Corollary}

\theoremstyle{definition}
\newtheorem{definition}[theorem]{Definition}

\theoremstyle{remark}
\newtheorem{remark}[theorem]{Remark}

\AtEndEnvironment{theorem}{\statementqed}
\AtEndEnvironment{lemma}{\statementqed}
\AtEndEnvironment{proposition}{\statementqed}
\AtEndEnvironment{corollary}{\statementqed}
\AtEndEnvironment{definition}{\statementqed}
\AtEndEnvironment{remark}{\statementqed}
\AtEndEnvironment{example}{\statementqed}

\newcommand{\pvec}[1]{\boldsymbol{\operatorname{#1}}}

\title{Quantum $f$-divergences from retrodictive tests: a~unifying classical variational principle}
\author{Francesco Buscemi}
\email{buscemi@nagoya-u.jp}
\affiliation{Dept. Mathematical Informatics, Nagoya University, Nagoya 464-8601, Japan}
\date{26 October 2026}

\begin{document}
\maketitle

\begin{abstract}
A single classical $f$-divergence has several inequivalent quantum versions, but what distinguishes them operationally is less clear. Matsumoto's reverse tests characterize the maximal quantum $f$-divergence as the least classical distinguishability needed to prepare two quantum states with one device. Here we introduce \textit{retrodictive tests}, in which two reverse tests sharing a reference state are coupled via Bayesian retrodiction to define a forward and a reverse stochastic process. In the forward process, the system is prepared by the first device and measured by the Petz transpose of the second; the reverse process interchanges their roles. We seek the least classical distinguishability between these two processes, as measured by the $f$-divergence between their joint laws, for fixed prescribed states and reference. For full-rank states in finite dimension, we prove that the canonical spectral implementations attain this minimum simultaneously for every operator-convex $f$. The reference then determines which quantum $f$-divergence is obtained: a flat reference gives the standard divergences, with the Nussbaum--Szko\l{}a distributions as optimal joint laws; either prescribed state as reference recovers Matsumoto's reverse tests and the maximal divergences; and power references give the geodesic family. In particular, the Umegaki and Belavkin--Staszewski relative entropies arise from the same classical variational principle under different choices of reference. Spectral optimality extends to processes containing a quantum channel; for channels with full-rank Choi operator, the minimum of the forward relative entropy agrees with a known average entropy production.
\end{abstract}

\section{Introduction}\label{sec:operational-question}

Statistical distinguishability is determined by the probability laws of experimental records, and therefore depends on how the system is prepared, how it is measured, and which information is retained. For quantum states, moreover, the same classical divergence admits inequivalent extensions to density operators, agreeing on commuting states but generally differing otherwise. The question is therefore not only how to define these extensions, but also which operational choices account for their differences.

Matsumoto's reverse tests provide one such account \citep{matsumoto2010reverse,matsumoto2018fdivergence}. A reverse test uses a single classical-to-quantum preparation channel with two input probability distributions, whose average outputs are the two prescribed quantum states. For operator-convex $f$, minimizing the classical $f$-divergence between these input distributions over all such reverse tests gives the corresponding \textit{maximal} quantum $f$-divergence between the prescribed outputs. Thus this construction interprets maximal quantum divergences as the least classical distinguishability needed to prepare both states through the same device.

Can an analogous interpretation be given also to other, non-maximal quantum divergences? Indeed, the standard quantum $f$-divergence \citep{petz1986quasi} also admits a classical representation through the Nussbaum--Szko\l{}a distributions \citep{nussbaum2009chernoff,hiai2011quantum,androulakis2024quantum}. This suggests seeking a generalization of reverse tests in which these distributions arise as minimizers of classical distinguishability.

To address this question, we consider \textit{two} reverse tests sharing a common reference output. One preparation device can produce either a prescribed state \(\rho\) or the reference \(\gamma\), while another can produce either a prescribed state \(\sigma\) or the same reference. These devices need not have the same classical input alphabet, and their separate descriptions do not yet specify how their labels should be coupled.

To couple the tests, we use Bayesian retrodiction, namely inference about the input of a process from its output relative to a prior. We adopt Petz transposition as the quantum inference rule, as motivated by structural axioms for retrodiction \citep{parzygnat2023retrodiction,surace2026inevitability} and by the \textit{quantum minimum-change principle} \citep{bai2025minimumchange,kuang2026measurement}. For each preparation channel, the classical input distribution that produces \(\gamma\) serves as the reference prior for Petz transposition, which determines the associated retrodictive measurement.

In the forward process ($F$), the first device is supplied with the input distribution that prepares \(\rho\), and the system is measured by the Petz transpose of the second device; the reverse process ($R$) interchanges their roles, starting from the input distribution that prepares \(\sigma\). Retaining both classical labels gives joint laws \(F\) and \(R\) on a common alphabet. Notice that, at the respective reference input distributions, the two processes have the same joint law; their conditional transitions are therefore Bayesian inverses with respect to these distributions. We call this coupled experiment a \textit{retrodictive test}.

For fixed full-rank states \(\rho,\sigma,\gamma\) in finite dimension, we minimize the classical divergence \(D_f(R\Vert F)\) between the reverse and forward joint laws over all admissible implementations. The canonical spectral reverse tests for \((\rho,\gamma)\) and \((\sigma,\gamma)\) solve this optimization simultaneously for every operator-convex \(f\). Their separate optimality does not by itself imply this result, since changing either preparation device also changes the coupling. The proof shows that replacing either test by its spectral realization cannot increase the coupled divergence, whatever the choice of the other test.

For an arbitrary reference, the minimum is a Petz quasi-entropy \citep{petz1986quasi}; it compares the complete processes and can remain positive even when \(\rho=\sigma\). A flat reference gives the standard quantum $f$-divergence and makes the Nussbaum--Szko\l{}a distributions optimal joint laws of the experiment. At either \textit{endpoint reference}, \(\gamma\in\{\rho,\sigma\}\), one of the two canonical tests is trivial, and the optimization reduces to Matsumoto's reverse-test problem. Power references connect the flat reference to \(\sigma\), recovering the geodesic quantum $f$-divergences \citep{capel2026geodesic}. In particular, for \(f(u)=u\log u\), changing the reference gives the passage from Umegaki to Belavkin--Staszewski relative entropy while leaving the classical criterion of distinguishability unchanged.

Sections~\ref{sec:admissible-implementations}--\ref{sec:reference-choices} develop the experiment, prove spectral optimality, and identify the divergences selected by the reference. In Section~\ref{sec:mixing-implementations}, mixing admissible implementations gives operational proofs of separate convexity and a stronger logarithmic mixture bound for relative entropy. Section~\ref{sec:channel-extension} inserts a quantum channel between preparation and measurement, with its Petz transpose in the reverse process. Spectral optimality remains valid, and for channels with full-rank Choi operator the minimum of the forward relative entropy \(D(F\Vert R)\) agrees with the average entropy production introduced in \citep{bai2026entropy}.

\section{Retrodictive tests from two reverse tests}\label{sec:admissible-implementations}

A \textit{reverse test} for an ordered pair of quantum states consists of a classical-to-quantum preparation channel and two probability distributions on its finite input alphabet, which the channel transforms to the respective states \citep{matsumoto2010reverse,matsumoto2018fdivergence}.

Let us now fix three full-rank (or \textit{faithful}) density operators \(\rho,\sigma,\gamma\) on a finite-dimensional Hilbert space \(\mathcal H\). We consider a reverse test for \((\rho,\gamma)\) and another for \((\sigma,\gamma)\), with preparation channels
\begin{align*}
\mathcal C(\pvec{x})&=\sum_{i\in\mathcal{I}} x_i\tau_i,
\qquad
\mathcal D(\pvec{y})=\sum_{j\in\mathcal{J}} y_j\upsilon_j,
\end{align*}
where the finite input alphabets $\mathcal{I}$ and $\mathcal{J}$ may differ and each \(\tau_i,\upsilon_j\) is a density operator.

\begin{definition}\label{def:admissible-implementation}
An \textit{admissible implementation} consists of two preparation channels \(\mathcal C,\mathcal D\), strictly positive reference probability distributions \(\pvec{g}=\{g_i\}\) and \(\pvec{h}=\{h_j\}\), and probability distributions \(\pvec{a}=\{a_i\}\) and \(\pvec{b}=\{b_j\}\), which may have zero entries, such that
\begin{align}
\mathcal C(\pvec{a})&=\rho,
&\mathcal D(\pvec{b})&=\sigma,\notag\\
\mathcal C(\pvec{g})&=\gamma,
&\mathcal D(\pvec{h})&=\gamma.
\label{eq:ensemble-constraints}
\end{align}
Thus \((\mathcal C,\pvec{a},\pvec{g})\) and \((\mathcal D,\pvec{b},\pvec{h})\) are reverse tests for \((\rho,\gamma)\) and \((\sigma,\gamma)\), respectively.
\end{definition}

\subsection{Coupling the tests by retrodiction}

To couple the two reverse tests, define the associated POVMs
\begin{align}
E_i&=g_i\gamma^{-1/2}\tau_i\gamma^{-1/2},
\qquad
H_j=h_j\gamma^{-1/2}\upsilon_j\gamma^{-1/2}.
\label{eq:petz-measurements}
\end{align}
Indeed, the effects are positive and sum to the identity. These measurement channels are the Petz transposes of \(\mathcal C\) and \(\mathcal D\) computed with respect to the classical reference laws \(\pvec{g}\) and \(\pvec{h}\), respectively. Conversely, Petz-transposing either measurement at \(\gamma\) recovers the corresponding preparation channel. Alternatively, we can think of
$\tau_i=\sqrt\gamma E_i\sqrt\gamma/g_i$
as the state \textit{retrodicted} from outcome $i$
when the prior is $\gamma$
\citep{bai2025minimumchange,kuang2026measurement},
according to the Petz rule, whose uniqueness as a
general retrodiction assignment under suitable
structural requirements was recently established
in Ref.~\citep{surace2026inevitability}.

\begin{figure}[t]
\centering
\includegraphics[width=\textwidth]{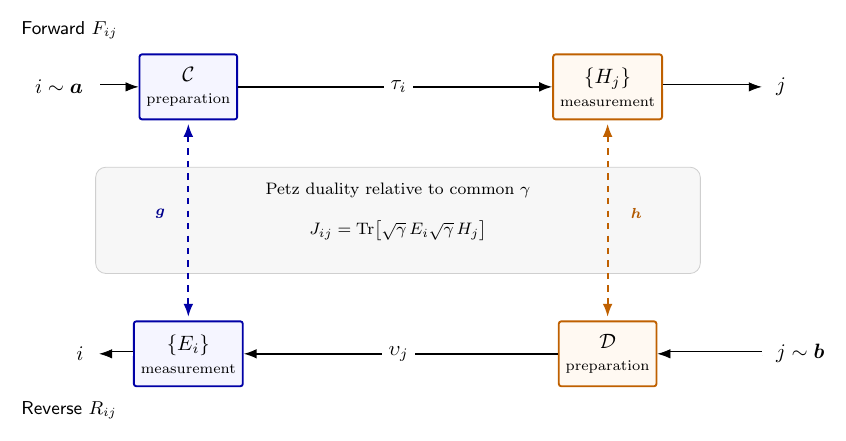}
\caption{Forward (upper) and reverse (lower) processes of a retrodictive test. The dashed vertical arrows indicate Petz transposition between each preparation channel and its corresponding measurement, relative to the common quantum reference $\gamma$ and the respective classical reference distributions $\boldsymbol{g}$ and $\boldsymbol{h}$. These arrows represent duality relations, not physical connections between the circuits. At the reference distributions, the two processes induce the same joint law $J_{ij}$ (central panel). In both processes, the complete classical record is the ordered pair $(i,j)$.}
\label{fig:admissible-processes}
\end{figure}

A \textit{retrodictive test} for \((\rho,\sigma)\) relative to \(\gamma\) couples the two reverse tests in an admissible implementation by measuring the output of each preparation channel with the Petz transpose of the other. The resulting forward and reverse processes are specified below and in Figure~\ref{fig:admissible-processes}.

\begin{center}
\renewcommand{\arraystretch}{1.2}
\begin{tabularx}{\textwidth}{@{}>{\raggedright\arraybackslash}p{0.26\textwidth}XX@{}}
\toprule
Operation & Forward process & Reverse process \\
\midrule
Draw initial label & $i$ from $\pvec{a}$ & $j$ from $\pvec{b}$ \\
Prepare the system & $\tau_i$ through $\mathcal C$ & $\upsilon_j$ through $\mathcal D$ \\
Measure the system & $\{H_j\}$, obtaining $j$ & $\{E_i\}$, obtaining $i$ \\
Retain both labels & $(i,j)$ & $(i,j)$ \\
\bottomrule
\end{tabularx}
\end{center}

Their joint laws are
\begin{align}
F_{ij}&\equiv a_iT(j\mid i):= a_i\operatorname{Tr}[\tau_iH_j],
\qquad
R_{ij}\equiv b_jT^\leftarrow(i\mid j):= b_j\operatorname{Tr}[\upsilon_jE_i].
\label{eq:complete-record-laws}
\end{align}
The outcome marginals are
\begin{align}
\sum_iF_{ij}&=\operatorname{Tr}[\rho H_j],
\qquad
\sum_jR_{ij}=\operatorname{Tr}[\sigma E_i].
\label{eq:record-marginals}
\end{align}
At the reference input distributions, both processes have the same joint law,
\begin{align*}
J_{ij}&:=g_iT(j\mid i)=h_jT^\leftarrow(i\mid j)
=\operatorname{Tr}[\sqrt\gamma E_i\sqrt\gamma H_j].
\end{align*}
The law \(J\), fixed by the two POVMs and the reference, has marginals \(\pvec{g},\pvec{h}\), so \(T^\leftarrow\) is the Bayesian inverse of \(T\) at \(\pvec{g}\). This is the classical consequence of reversal of composition under Petz transposition \citep{parzygnat2023retrodiction,surace2026inevitability}, also used in Bayesian fluctuation relations \citep{buscemi2021fluctuation,aw-buscemi-scarani}. The actual input distributions give
\begin{align}
F_{ij}&=\lambda_iJ_{ij},
\qquad
R_{ij}=\mu_jJ_{ij},
\end{align}
where $\lambda_i:=a_i/g_i$ and $\mu_j:=b_j/h_j$ are the classical likelihood ratios of the two reverse tests.

\subsection{Weighted POVM parametrization}

Admissibility can be expressed directly in terms of the retrodictive POVMs and their likelihood ratios. Define the relative density operators
\begin{align}
A&=\gamma^{-1/2}\rho\gamma^{-1/2},
\qquad
B=\gamma^{-1/2}\sigma\gamma^{-1/2}.
\label{eq:relative-densities}
\end{align}
Equations~\eqref{eq:ensemble-constraints} and~\eqref{eq:petz-measurements} then give
\begin{align}
\sum_i\lambda_iE_i=A,
\qquad
\sum_j\mu_jH_j=B.
\label{eq:weighted-povms}
\end{align}
Conversely, any two finite POVMs with nonzero effects and nonnegative weights satisfying \eqref{eq:weighted-povms} determine an admissible implementation\footnote{Given any finite decomposition \(A=\sum_iD_i\) into nonzero positive semidefinite operators, choose \(\lambda_i>0\) such that \(\sum_iD_i/\lambda_i\leq I\); taking every \(\lambda_i\geq\|A\|_\infty\) suffices. Then \(E_i=D_i/\lambda_i\), together with \(E_0=I-\sum_iE_i\) and \(\lambda_0=0\), defines a POVM with weighted sum \(A\). Omit \(E_0\) if it vanishes. The same construction applies to \(B\).} through
\begin{align}
g_i&=\operatorname{Tr}[\gamma E_i],&
\tau_i&=\frac{\sqrt\gamma E_i\sqrt\gamma}{g_i},&
a_i&=\lambda_i g_i,\notag\\
h_j&=\operatorname{Tr}[\gamma H_j],&
\upsilon_j&=\frac{\sqrt\gamma H_j\sqrt\gamma}{h_j},&
b_j&=\mu_j h_j.
\label{eq:retrodictive-preparations}
\end{align}
Indeed, faithfulness of \(\gamma\) ensures \(g_i,h_j>0\), while POVM normalization and \eqref{eq:weighted-povms} give the required ensemble averages and normalized probability distributions.

\section{Spectral optimality}\label{sec:spectral-optimality}

We now specify the criterion used to compare the two joint laws. For a convex generator \(f:(0,\infty)\to\mathbb R\) with \(f(1)=0\), the classical \(f\)-divergence is
\begin{align}
D_f(R\Vert F)
&=\sum_{i,j:F_{ij}>0}F_{ij}f\!\left(\frac{R_{ij}}{F_{ij}}\right)
+f'_\infty\sum_{i,j:F_{ij}=0}R_{ij},
\qquad
f'_\infty=\lim_{u\to\infty}\frac{f(u)}{u}.
\label{eq:classical-f-divergence}
\end{align}
At the boundary, \(f(0)=\lim_{u\downarrow0}f(u)\), and a record with \(F_{ij}=R_{ij}=0\) contributes zero, including when \(f'_\infty=+\infty\).

The distinction from two separate reverse-test optimizations is explicit. Since \(J\) has marginals \(\pvec{g},\pvec{h}\),
\begin{align*}
D_f(F\Vert J)&=D_f(\pvec{a}\Vert\pvec{g}),
&D_f(R\Vert J)&=D_f(\pvec{b}\Vert\pvec{h}).
\end{align*}
Thus each separate reverse-test cost compares one joint law with \(J\), whereas our objective compares \(R\) and \(F\) directly. Its value depends on how the likelihood ratios \(\lambda_i,\mu_j\) are paired through \(J\), and cannot in general be recovered from the two separate divergence values. To solve the coupled optimization, we restrict to operator-convex generators \(f\), so that operator Jensen applies to the weighted POVM resolutions in \eqref{eq:weighted-povms}.

For \(f(u)=u\log u\), the classical divergence is the relative entropy (Kullback--Leibler divergence),
\begin{align*}
D(R\Vert F)&=\sum_{i,j:R_{ij}>0}R_{ij}\log\frac{R_{ij}}{F_{ij}},
\end{align*}
with natural logarithms and value \(+\infty\) if \(R_{ij}>0=F_{ij}\) for some record. The conjugate generator \(f(u)=-\log u\) gives the opposite orientation, \(D(F\Vert R)\).

\begin{theorem}[Simultaneous spectral optimality]\label{thm:spectral-optimality}
Let \(\rho,\sigma,\gamma>0\) be density operators on the same finite-dimensional Hilbert space, and let \(f:(0,\infty)\to\mathbb R\) be operator convex, with \(f(1)=0\). Define \(A=\gamma^{-1/2}\rho\gamma^{-1/2}\) and \(B=\gamma^{-1/2}\sigma\gamma^{-1/2}\), and write their spectral decompositions
\begin{align}
A&=\sum_k\alpha_kP_k,
\qquad
B=\sum_l\beta_lQ_l,
\label{eq:spectral-decompositions}
\end{align}
where the eigenvalues are distinct and the projections are the full, possibly higher-rank spectral projections. Over all finite admissible implementations of Section~\ref{sec:admissible-implementations},
\begin{align}
\min D_f(R\Vert F)
&=\sum_{k,l}\alpha_k f\!\left(\frac{\beta_l}{\alpha_k}\right)
\operatorname{Tr}[\sqrt\gamma P_k\sqrt\gamma Q_l]=: V_f(\rho,\sigma;\gamma).
\label{eq:spectral-minimum}
\end{align}
The spectral POVMs \(\{P_k\}\) and \(\{Q_l\}\), with weights \(\alpha_k,\beta_l\), attain this minimum simultaneously for every such \(f\).
\end{theorem}

\begin{proof}
Let \(\{E_i\}\) and \(\{H_j\}\), with weights \(\lambda_i,\mu_j\), be arbitrary admissible resolutions of \(A,B\). The proof applies operator Jensen twice, replacing one resolution at a time by its spectral decomposition.

First suppose that all \(\lambda_i,\mu_j\) are strictly positive. Using \eqref{eq:complete-record-laws}, with records for which \(J_{ij}=0\) contributing zero, gives
\begin{align*}
D_f(R\Vert F)
&=\sum_{i,j}\lambda_i f\!\left(\frac{\mu_j}{\lambda_i}\right)
\operatorname{Tr}[\sqrt\gamma E_i\sqrt\gamma H_j]\\
&=\sum_i\lambda_i\operatorname{Tr}\!\left[
\sqrt\gamma E_i\sqrt\gamma
\sum_j f\!\left(\frac{\mu_j}{\lambda_i}\right)H_j
\right].
\end{align*}
For each fixed \(i\), operator Jensen \citep{hansen2003jensen}, applied to the POVM \(\{H_j\}\), gives
\begin{align*}
\sum_j f\!\left(\frac{\mu_j}{\lambda_i}\right)H_j
&\geq f\!\left(\sum_j\frac{\mu_j}{\lambda_i}H_j\right)
=f(B/\lambda_i).
\end{align*}
Taking the trace against the positive operator \(\lambda_i\sqrt\gamma E_i\sqrt\gamma\) preserves this inequality. Summing over \(i\), expanding \(B\) spectrally, and using cyclicity of the trace, we obtain
\begin{align}
D_f(R\Vert F)
&\geq\sum_i\lambda_i\operatorname{Tr}[\sqrt\gamma E_i\sqrt\gamma f(B/\lambda_i)]\notag\\
&=\sum_{i,l}\lambda_i f(\beta_l/\lambda_i)
\operatorname{Tr}[\sqrt\gamma E_i\sqrt\gamma Q_l]\notag\\
&=\sum_l\operatorname{Tr}\!\left[
\sqrt\gamma Q_l\sqrt\gamma
\sum_i\lambda_i f(\beta_l/\lambda_i)E_i
\right].
\label{eq:first-jensen}
\end{align}
Thus replacing the weighted POVM resolution of \(B\) by its spectral resolution cannot increase the divergence, while the resolution of \(A\) is still arbitrary.

For the second resolution, use the conjugate generator \(\widetilde f(u):=u f(1/u)\), which is operator convex whenever \(f\) is \citep{matsumoto2018fdivergence}. For each \(\beta_l>0\), the function \(u\mapsto u f(\beta_l/u)=\beta_l\widetilde f(u/\beta_l)\) is therefore operator convex. Applying operator Jensen to \(\{E_i\}\) gives
\begin{align}
\sum_i\lambda_i f(\beta_l/\lambda_i)E_i
&\geq A f(\beta_l A^{-1}).
\label{eq:second-jensen}
\end{align}
Since \(\sqrt\gamma Q_l\sqrt\gamma\geq0\), we can insert this inequality into \eqref{eq:first-jensen}. Expanding \(A\) spectrally then yields
\begin{align*}
D_f(R\Vert F)
&\geq\sum_l\operatorname{Tr}[\sqrt\gamma Q_l\sqrt\gamma A f(\beta_l A^{-1})]\\
&=\sum_{k,l}\alpha_k f\!\left(\frac{\beta_l}{\alpha_k}\right)
\operatorname{Tr}[\sqrt\gamma P_k\sqrt\gamma Q_l],
\end{align*}
which is the required lower bound.

In the case in which some of the weights vanish, replace \(\lambda_i,\mu_j\) by \(\lambda_i+\varepsilon,\mu_j+\varepsilon\) in the same two Jensen inequalities. The corresponding weighted sums are \(A+\varepsilon I,B+\varepsilon I\), with the same spectral projections as \(A,B\). As \(\varepsilon\downarrow0\), each summand
\begin{align*}
(\lambda_i+\varepsilon)
f\!\left(\frac{\mu_j+\varepsilon}{\lambda_i+\varepsilon}\right)J_{ij}
\end{align*}
converges to its extended value in \eqref{eq:classical-f-divergence}. In particular, when \(\lambda_i=\mu_j=0\), this summand is identically zero because \(f(1)=0\). The spectral expression has a finite limit because \(A,B>0\). Hence the same lower bound holds for every admissible implementation.

Finally, choose \(E_k=P_k\), \(\lambda_k=\alpha_k\), \(H_l=Q_l\), and \(\mu_l=\beta_l\). These choices are admissible, and both Jensen inequalities become equalities by spectral functional calculus. The spectral implementation therefore attains the lower bound. Since this choice depends only on \(A,B\), and not on \(f\), it attains the minimum simultaneously for every generator in the stated class.
\end{proof}

Since \(D_f(R\Vert F)=D_{\widetilde f}(F\Vert R)\), the first Jensen step also applies with the two resolutions interchanged. Thus either reverse test can be replaced by its canonical spectral realization while the other remains arbitrary; this is the additional property that allows the two separate reverse-test optimizers to solve the coupled problem.

\begin{remark}[Identification with a Petz quasi-entropy]\label{rem:quasi-entropy}
For an arbitrary reference, the minimum in \eqref{eq:spectral-minimum} is a \textit{Petz quasi-entropy} \citep{petz1986quasi}. To see this, recall that for a fixed operator \(K\), its spectral expression is
\begin{align*}
S_f^K(B\Vert A)
&=\sum_{k,l}\alpha_k f\!\left(\frac{\beta_l}{\alpha_k}\right)
\operatorname{Tr}[K^\dagger Q_l K P_k].
\end{align*}
Taking \(K=\sqrt\gamma\) and using cyclicity of the trace therefore gives
\begin{align*}
V_f(\rho,\sigma;\gamma)=S_f^{\sqrt\gamma}(B\Vert A).
\end{align*}
\end{remark}

\begin{corollary}[Minimum classical relative entropy]
For faithful \(\rho,\sigma,\gamma\) on the same finite-dimensional Hilbert space, the minimum of the relative entropy (Kullback--Leibler divergence) over the admissible implementations of Section~\ref{sec:admissible-implementations} is
\begin{align}
\min D(R\Vert F)
&=\operatorname{Tr}[\gamma B\log B]
-\operatorname{Tr}[\sigma\log A]=:V(\rho,\sigma;\gamma).
\label{eq:relative-entropy-minimum}
\end{align}
\end{corollary}

\begin{proof}
Substitute \(f(u)=u\log u\) in \eqref{eq:spectral-minimum}. The first logarithmic term sums to \(\operatorname{Tr}[\gamma B\log B]\), since \(\sum_kP_k=I\). For the second, use \(\sqrt\gamma B\sqrt\gamma=\sigma\), which gives
\begin{align*}
\sum_{k,l}\beta_l\log\alpha_k\operatorname{Tr}[\sqrt\gamma P_k\sqrt\gamma Q_l]
&=\operatorname{Tr}[\sigma\log A].
\end{align*}
\end{proof}

\begin{remark}[Conditions for complete indistinguishability]\label{rem:reference-dependence}
For faithful \(\rho,\sigma,\gamma\), the minimum in \eqref{eq:relative-entropy-minimum} satisfies
\begin{align*}
V(\rho,\sigma;\gamma)=0
\quad\Longleftrightarrow\quad
\rho=\sigma\ \text{and}\ [\rho,\gamma]=0.
\end{align*}
Indeed, at the spectral optimizer, \(J_{kl}=0\) if and only if \(P_k\sqrt\gamma Q_l=0\). Hence equality of the joint laws is equivalent to \((\alpha_k-\beta_l)P_k\sqrt\gamma Q_l=0\) for every \(k,l\), or
\begin{align*}
A\sqrt\gamma&=\sqrt\gamma B.
\end{align*}
Substituting \eqref{eq:relative-densities} gives \(\rho=\sqrt\gamma\,\sigma\gamma^{-1/2}\). Taking adjoints gives also \(\rho=\gamma^{-1/2}\sigma\sqrt\gamma\); equating these expressions yields \([\sigma,\gamma]=0\), and therefore \(\rho=\sigma\). Conversely, these conditions give identical spectral joint laws and hence zero minimum.

Thus, if the prescribed states coincide but do not commute with the reference, every admissible implementation gives distinguishable forward and reverse processes.
\end{remark}

\section{Quantum divergences selected by the reference}\label{sec:reference-choices}

We now choose the reference as a prescribed function of \(\rho,\sigma\), and identify the resulting standard, maximal, and geodesic quantum $f$-divergences. For these choices, the reference commutes with the common state when \(\rho=\sigma\), so the minimum vanishes at equal arguments. Throughout this section, \(f:(0,\infty)\to\mathbb R\) is operator convex, with \(f(1)=0\).

\subsection{Flat reference: standard divergences and Nussbaum--Szko\l{}a distributions}\label{sec:flat-reference}

Let \(\rho,\sigma>0\) be states on a \(d\)-dimensional Hilbert space and take \(\gamma=I/d\). Write \(\rho=\sum_i r_iP_i\) and \(\sigma=\sum_j s_jQ_j\) with full spectral projections. Since \(A=d\rho\) and \(B=d\sigma\), \eqref{eq:spectral-minimum} gives
\begin{align*}
\left.\min D_f(R\Vert F)\right|_{\gamma=I/d}
&=\sum_{i,j}r_i f(s_j/r_i)\operatorname{Tr}[P_iQ_j]
=:D_f^{\mathrm{Petz}}(\sigma\Vert\rho).
\end{align*}
The expression on the right is the standard quantum \(f\)-divergence \citep{petz1986quasi}. The spectral preparations and joint laws attaining it are
\begin{align*}
\tau_i&=\frac{P_i}{\operatorname{rank}P_i},
&
\upsilon_j&=\frac{Q_j}{\operatorname{rank}Q_j},\\
F_{ij}&=r_i\operatorname{Tr}[P_iQ_j],
&
R_{ij}&=s_j\operatorname{Tr}[P_iQ_j].
\end{align*}
Resolving the eigenspaces into orthonormal bases gives the usual Nussbaum--Szko\l{}a distributions \citep{nussbaum2009chernoff,hiai2011quantum,androulakis2024quantum}, \(F_{ij}=r_i|\langle r_i\mid s_j\rangle|^2\) and \(R_{ij}=s_j|\langle r_i\mid s_j\rangle|^2\), with eigenvalues repeated according to multiplicity. On records of positive probability, the refinement leaves each likelihood ratio \(R_{ij}/F_{ij}=s_j/r_i\) unchanged, so it preserves every classical \(f\)-divergence under consideration.

In this refined realization, the forward process prepares an eigenstate of \(\rho\) and measures in an eigenbasis of \(\sigma\); the reverse process interchanges these roles. This realization is the identity-channel specialization of \citep{buscemi2021fluctuation}. The additional conclusion here is the optimality of these joint laws over all admissible implementations, simultaneously for every operator-convex generator \(f\).

In particular, choosing \(f(u)=u\log u\) gives the Umegaki relative entropy:
\begin{align*}
\left.\min D(R\Vert F)\right|_{\gamma=I/d}
&=\operatorname{Tr}[\sigma(\log\sigma-\log\rho)]
=D(\sigma\Vert\rho).
\end{align*}

\subsection{Endpoint references: maximal divergences and Matsumoto's reverse tests}\label{sec:endpoint-references}

At either endpoint reference, \(\gamma\in\{\rho,\sigma\}\), the coupled optimization reduces to Matsumoto's reverse-test problem. At \(\gamma=\rho\), one has \(A=I\), so the first spectral test has a single label and always prepares \(\rho\).

For any admissible implementation at this reference, the constraints on the reverse preparation ensemble and the marginal identities \eqref{eq:record-marginals} give
\begin{align*}
\rho&=\sum_jh_j\upsilon_j,
&
\sigma&=\sum_jb_j\upsilon_j,
\\
\sum_iF_{ij}&=h_j,
&
\sum_iR_{ij}&=b_j.
\end{align*}
Hence a device that, upon receiving the complete record \((i,j)\), ignores label $i$ and prepares \(\upsilon_j\), converts \(F\) into \(\rho\) and \(R\) into \(\sigma\). In other words, every admissible implementation at this reference defines a reverse test on its complete record alphabet. The same conclusion holds at \(\gamma=\sigma\), using a device that, upon receiving \((i,j)\), ignores $j$ and prepares \(\tau_i\).

Conversely, consider a reverse test consisting of a preparation channel \(\mathcal W(\pvec{x})=\sum_jx_j\omega_j\) and distributions satisfying \(\mathcal W(\pvec{p})=\rho\), \(\mathcal W(\pvec{q})=\sigma\). If every \(p_j>0\), choose the first test to be the one-label preparation of \(\rho\), and the second to be \((\mathcal W,\pvec{q},\pvec{p})\). At reference \(\rho\), this is an admissible implementation with joint laws \(F_{*j}=p_j\) and \(R_{*j}=q_j\). If every \(q_j>0\), the same reverse test embeds at reference \(\sigma\) by choosing the first test as \((\mathcal W,\pvec{p},\pvec{q})\) and the second as the one-label preparation of \(\sigma\); then \(F_{j*}=p_j\) and \(R_{j*}=q_j\).

\begin{figure}[t]
\centering
\includegraphics[width=\textwidth]{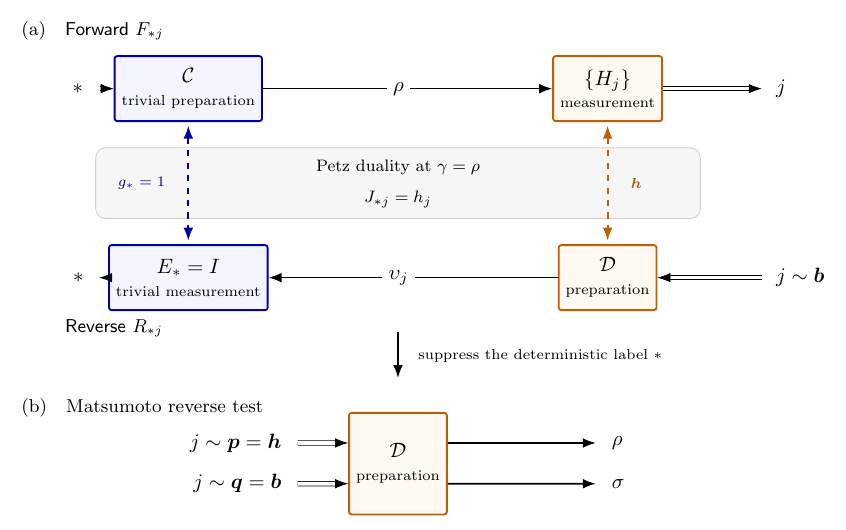}
\caption{Reduction to Matsumoto's reverse test.
(a) At the endpoint reference $\gamma=\rho$, choose
the first reverse test to be trivial:
$\mathcal C(*)=\rho$ and $E_*=I$, with
$a_*=g_*=1$. The second preparation channel
$\mathcal D$ is paired with its Petz transpose
$\{H_j\}$ relative to $(\pvec{h},\rho)$.
Consequently, $J_{*j}=F_{*j}=h_j$ and
$R_{*j}=b_j$.
(b) Suppressing the deterministic label $*$
leaves a single preparation channel $\mathcal D$
that produces $\rho$ from $\pvec{p}=\pvec{h}$
and $\sigma$ from $\pvec{q}=\pvec{b}$,
which is precisely a Matsumoto reverse test.
The two input rows represent alternative
uses of the same device.}
\label{fig:reverse-test-reduction}
\end{figure}

At \(\gamma=\rho\), write \(B=\rho^{-1/2}\sigma\rho^{-1/2}=\sum_j\mu_jQ_j\). Together with \(E_*=I\), the spectral choice \(H_j=Q_j\) gives the preparation data
\begin{align}
p_j&=\operatorname{Tr}[\rho Q_j],
\qquad
q_j=\mu_jp_j,
\qquad
\upsilon_j=\frac{\sqrt\rho Q_j\sqrt\rho}{p_j}.
\label{eq:reverse-test-construction}
\end{align}
These distributions and prepared states constitute the canonical Matsumoto reverse test \citep{matsumoto2010reverse,matsumoto2018fdivergence,lanier2025classical,ruibal2025unraveling}. Since both distributions are strictly positive, this same optimal test is available at either endpoint. Resolving degenerate eigenspaces into rank-one projections gives its usual pure-state form, without changing the likelihood ratios or any classical \(f\)-divergence. Figure~\ref{fig:reverse-test-reduction} illustrates the reduction from the two reference ensembles to this single preparation device.

The common minimum is therefore
\begin{align}
\left.\min D_f(R\Vert F)\right|_{\gamma=\rho}
&=\left.\min D_f(R\Vert F)\right|_{\gamma=\sigma}
\notag\\
&=\min_{\mathcal W,\pvec{p},\pvec{q}}
\left\{D_f(\pvec{q}\Vert\pvec{p}):
\mathcal W(\pvec{p})=\rho\ \wedge\
\mathcal W(\pvec{q})=\sigma\right\}
\notag\\
&=\operatorname{Tr}[\rho f(B)]
=:\widehat D_f(\sigma\Vert\rho).
\label{eq:maximal-minimum}
\end{align}
The quantity on the right is the maximal quantum \(f\)-divergence. The reverse-test minimum also includes boundary input distributions, with the conventions of \eqref{eq:classical-f-divergence}.\footnote{After removing labels with \(p_j=q_j=0\), put \(\pvec{p}^{\,\varepsilon}=(1-\varepsilon)\pvec{p}+\varepsilon\pvec{q}\) and \(\rho_\varepsilon=(1-\varepsilon)\rho+\varepsilon\sigma\). The construction above applies to the strictly positive law \(\pvec{p}^{\,\varepsilon}\), and the spectral lower bound at reference \(\rho_\varepsilon\) gives \(D_f(\pvec{q}\Vert\pvec{p}^{\,\varepsilon})\geq\widehat D_f(\sigma\Vert\rho_\varepsilon)\). Letting \(\varepsilon\downarrow0\) gives the same lower bound for the original test. The classical expression converges to its prescribed extended value, while the quantum expression is continuous because the states remain faithful. Thus boundary tests satisfy the same lower bound, without necessarily embedding at either fixed endpoint.}

At the other endpoint, \(\gamma=\sigma\), one has \(B=I\) and \(A=\sigma^{-1/2}\rho\sigma^{-1/2}\). Spectral optimality therefore also gives
\begin{align*}
\widehat D_f(\sigma\Vert\rho)
&=\operatorname{Tr}[\sigma A f(A^{-1})].
\end{align*}
For \(f(u)=u\log u\), the common minimum is the Belavkin--Staszewski relative entropy,\footnote{This minimum also admits a Bayesian interpretation. For a reverse test with finite \(D(\pvec{q}\Vert\pvec{p})\), remove labels with \(p_j=q_j=0\), choose \(c\geq\max_jq_j/p_j\), and take an evidence event \(\mathsf E\) with \(\Pr(\mathsf E\mid j)=q_j/(cp_j)\). Conditioning changes \(\pvec{p}\) into \(\pvec{q}\), while the prepared states \(\omega_j\) remain fixed. Thus the minimum is also the least classical information gain needed to explain the change from \(\rho\) to \(\sigma\) by learning about an unchanged preparation source.}
\begin{align}
D_{\mathrm{BS}}(\sigma\Vert\rho)
&=-\operatorname{Tr}[\sigma\log A]
=\operatorname{Tr}\!\left[
\sigma\log(\sqrt\sigma\,\rho^{-1}\sqrt\sigma)
\right].
\label{eq:bs-equivalent}
\end{align}

\subsection{\texorpdfstring{Power references: geodesic quantum \(f\)-divergences}{Power references: geodesic quantum f-divergences}}\label{sec:power-references}

To connect the flat reference to \(\sigma\), keep the prescribed states \(\rho,\sigma>0\) and the generator \(f\) fixed, and choose
\begin{align}
\gamma_t&=\frac{\sigma^t}{Z_t},
\qquad
Z_t=\operatorname{Tr}[\sigma^t],
\qquad
0\leq t\leq1.
\label{eq:power-reference}
\end{align}
The above choice interpolates between \(\gamma_0=I/d\) and \(\gamma_1=\sigma\); for each \(t\), the chosen reference is held fixed throughout the minimization.

The relative density operators are \(A_t=Z_tC_t\) and \(B_t=Z_t\sigma^{1-t}\), where \(C_t=\sigma^{-t/2}\rho\sigma^{-t/2}\). Write
\begin{align*}
C_t&=\sum_kc_kP_k,
\qquad
\sigma=\sum_ls_lQ_l,
\end{align*}
with full spectral projections; the dependence of \(c_k,P_k\) on \(t\) is suppressed in the notation. The spectral construction chooses \(\{Q_l\}\) as the forward measurement and \(\{P_k\}\) as the reverse measurement. Their corresponding preparations and preparation probabilities are
\begin{align*}
\tau_k(t)&=\frac{\sigma^{t/2}P_k\sigma^{t/2}}
{\operatorname{Tr}[\sigma^tP_k]},
&
a_k&=c_k\operatorname{Tr}[\sigma^tP_k],
\\
\upsilon_l&=\frac{Q_l}{\operatorname{rank}Q_l},
&
b_l&=s_l\operatorname{rank}Q_l.
\end{align*}
In particular, the forward measurement and the reverse preparation ensemble can be kept fixed throughout the family. Changing the reference changes \(C_t\), and hence the reverse measurement and the forward preparations.

The forward and reverse joint laws are
\begin{align}
F_t(k,l)&=c_ks_l^t\operatorname{Tr}[P_kQ_l],
\qquad
R_t(k,l)=s_l\operatorname{Tr}[P_kQ_l].
\label{eq:power-records}
\end{align}
At \(t=0\), these are the Nussbaum--Szko\l{}a distributions of Section~\ref{sec:flat-reference}. At \(t=1\), one has \(B_1=I\), and retaining \(\{Q_l\}\) merely refines its trivial spectral measurement: the likelihood ratio is \(1/c_k\) on the common support, independently of \(l\). Summing over \(l\) therefore preserves every classical \(f\)-divergence and gives the endpoint reverse test of Section~\ref{sec:endpoint-references}.

\begin{corollary}[Geodesic quantum \(f\)-divergences as minima]
Let \(\rho,\sigma>0\) be density operators on the same finite-dimensional Hilbert space, let \(0\leq t\leq1\), and let \(f\) satisfy the standing assumptions of this section. For the reference \(\gamma_t\) in \eqref{eq:power-reference},
\begin{align}
\left.\min D_f(R\Vert F)\right|_{\gamma=\gamma_t}
&=D_f^t(\sigma\Vert\rho),
\label{eq:geodesic-minimum}
\end{align}
where \(D_f^t\) is the geodesic quantum \(f\)-divergence of Capel and Costa Rico \citep{capel2026geodesic,capel2026geometry}. The minimum ranges over all finite admissible implementations and is attained by the joint laws \eqref{eq:power-records}.
\end{corollary}

\begin{proof}
With the state arguments ordered as here, the geodesic divergence is defined by
\begin{align}
D_f^t(\sigma\Vert\rho)
&=\left\langle\sqrt\rho,\,
f(\Gamma_t)(\sqrt\rho)\right\rangle_{\mathrm{HS}},
\notag\\
\Gamma_t(X)
&=\sigma^{1-t}X\rho^{-1/2}\sigma^t\rho^{-1/2},
\label{eq:geodesic-definition}
\end{align}
where \(\langle X,Y\rangle_{\mathrm{HS}}=\operatorname{Tr}[X^\dagger Y]\). To identify this expression with the divergence of our joint laws, set \(U_t=\rho^{-1/2}\sigma^{t/2}C_t^{1/2}\). Since \(\sigma^{t/2}C_t\sigma^{t/2}=\rho\), the operator \(U_t\) is unitary and satisfies
\begin{align*}
\rho^{-1/2}\sigma^t\rho^{-1/2}
&=U_tC_t^{-1}U_t^\dagger.
\end{align*}
Consequently, the mutually orthogonal projections \(X\mapsto Q_lXU_tP_kU_t^\dagger\) resolve \(\Gamma_t\), with eigenvalues \(s_l^{1-t}/c_k\). Their weights in \(\sqrt\rho\) are
\begin{align*}
\left\|Q_l\sqrt\rho\,U_tP_k\right\|_{\mathrm{HS}}^2
&=c_ks_l^t\operatorname{Tr}[P_kQ_l]
=F_t(k,l),
\end{align*}
where we used \(\sqrt\rho\,U_t=\sigma^{t/2}C_t^{1/2}\). Hence
\begin{align*}
D_f^t(\sigma\Vert\rho)
&=\sum_{k,l}F_t(k,l)
f\!\left(\frac{s_l^{1-t}}{c_k}\right)
=D_f(R_t\Vert F_t).
\end{align*}
Spectral optimality of the joint laws \eqref{eq:power-records} proves \eqref{eq:geodesic-minimum}.
\end{proof}

For \(f(u)=u\log u\), the direct substitution of \(A_t,B_t\) into \eqref{eq:relative-entropy-minimum} cancels the normalization terms \(\log Z_t\) and immediately gives
\begin{align}
\left.\min D(R\Vert F)\right|_{\gamma=\gamma_t}
&=D_t(\sigma\Vert\rho)
\notag\\
&=(1-t)\operatorname{Tr}[\sigma\log\sigma]
-\operatorname{Tr}\!\left[
\sigma\log(\sigma^{-t/2}\rho\sigma^{-t/2})
\right].
\label{eq:geodesic-relative-entropy}
\end{align}
The geometric interpretation and other properties of this family are studied in \citep{capel2026geodesic,capel2026geometry}.

\section{Convexity from mixing admissible implementations}\label{sec:mixing-implementations}

The mixture inequalities below also follow from known operator results; here we derive them by combining admissible preparation devices. If one prescribed state is fixed, its canonical reverse test can be used for every component, while the other tests are combined to prepare the varying state. We show that this gives separate convexity at fixed \(\gamma\); for relative entropy, allowing different component probabilities in the actual and reference preparations gives a stronger logarithmic bound.

\begin{proposition}[Separate convexity and a relative-entropy strengthening]
\label{prop:mixture-bounds}
Let \(\gamma>0\) be a fixed density operator, and let \(f:(0,\infty)\to\mathbb R\) be operator convex, with \(f(1)=0\). For every probability distribution \(\{p_k\}_{k=1}^m\), the following statements hold.
\begin{enumerate}
\item \emph{Separate convexity.}
For faithful density operators \(\rho,\sigma,\rho_k,\sigma_k\),
\begin{align*}
V_f\!\left(\sum_kp_k\rho_k,\sigma;\gamma\right)
&\leq \sum_kp_kV_f(\rho_k,\sigma;\gamma),\\
V_f\!\left(\rho,\sum_kp_k\sigma_k;\gamma\right)
&\leq \sum_kp_kV_f(\rho,\sigma_k;\gamma).
\end{align*}

\item \emph{Relative-entropy strengthening.}
In the relative-entropy case, for faithful density operators \(\sigma,\rho_1,\ldots,\rho_m\),
\begin{align}
V\!\left(\sum_kp_k\rho_k,\sigma;\gamma\right)
&\leq -\log\sum_kp_k
       \exp[-V(\rho_k,\sigma;\gamma)]\notag\\
&\leq \sum_kp_kV(\rho_k,\sigma;\gamma).
\label{eq:relative-entropy-mixture}
\end{align}
The second inequality is the scalar Jensen inequality; hence, the first inequality strengthens convexity in \(\rho\). Equivalently, the map $\rho\longmapsto \exp[-V(\rho,\sigma;\gamma)]$ is concave for fixed \(\sigma,\gamma\).
\end{enumerate}
\end{proposition}

\begin{proof}[Proof of Proposition~\ref{prop:mixture-bounds}]
Discard terms with \(p_k=0\). We combine optimal implementations, retaining the component label \(k\) so that both joint laws are defined on triples \((k,i,j)\). For each \(k\), choose a spectral implementation for \((\rho_k,\sigma,\gamma)\), with first weighted POVM resolution \(\{E_i^{(k)},\lambda_i^{(k)}\}\) and joint laws \(F^{(k)},R^{(k)}\). Since \(\sigma\) is fixed, the same spectral resolution \(\{H_j,\mu_j\}\) of \(B=\gamma^{-1/2}\sigma\gamma^{-1/2}\) can be used throughout.

In the forward process, we realize the prescribed mixture by choosing the component label \(k\) with distribution \(\pvec{p}=\{p_k\}\). In the reverse process, however, \(k\) can label a random choice of measurement, whose probabilities need not equal \(p_k\). For any strictly positive probability distribution \(\pvec{q}=\{q_k\}\), set
\[
\overline E_{k,i}=q_kE_i^{(k)},
\qquad
\overline\lambda_{k,i}=\frac{p_k}{q_k}\lambda_i^{(k)}.
\]
Since \(\sum_i\overline E_{k,i}=q_kI\), the reverse measurement chooses component \(k\) with probability \(q_k\), independently of the input state. The effects sum to \(I\), while \(\overline\lambda_{k,i}\overline E_{k,i} =p_k\lambda_i^{(k)}E_i^{(k)}\). Hence, together with the unchanged resolution of \(B\), they give an admissible implementation with prescribed averages \(\sum_kp_k\rho_k\) and \(\sigma\), and reference \(\gamma\). Equation~\eqref{eq:retrodictive-preparations} gives
\begin{align*}
\overline g_{k,i}&=q_kg_i^{(k)},&
\overline\tau_{k,i}&=\tau_i^{(k)},&
\overline a_{k,i}&=p_ka_i^{(k)}.
\end{align*}
The resulting joint laws are
\begin{align}
\overline F_{k,i,j}=p_kF^{(k)}_{ij},
\qquad
\overline R_{k,i,j}=q_kR^{(k)}_{ij}.
\label{eq:mixed-records}
\end{align}

Taking \(\pvec{q}=\pvec{p}\), admissibility and optimality of the component implementations give
\[
V_f\!\left(\sum_kp_k\rho_k,\sigma;\gamma\right)
\leq D_f(\overline R\Vert\overline F)
=\sum_kp_kV_f(\rho_k,\sigma;\gamma).
\]
Convexity in \(\sigma\) follows in the same way: keep the spectral resolution of \(A=\gamma^{-1/2}\rho\gamma^{-1/2}\) fixed and combine the second resolutions using \(\overline H_{k,j}=p_kH_j^{(k)}\) and \(\overline\mu_{k,j}=\mu_j^{(k)}\). Both joint laws then carry the same weights \(p_k\).

In the case of relative entropy, however, the freedom in \(\pvec{q}\) can be exploited to obtain a stronger bound. Writing \(v_k:=V(\rho_k,\sigma;\gamma)\), the classical chain rule yields
\begin{align}
V\!\left(\sum_kp_k\rho_k,\sigma;\gamma\right)
&\leq D(\overline R\Vert\overline F)\notag\\
&=D(\pvec{q}\Vert\pvec{p})+\sum_kq_kv_k.
\label{eq:mixture-trial-cost}
\end{align}
In particular, choosing
\[
q_k=\frac{p_ke^{-v_k}}{Z},
\qquad
Z:=\sum_\ell p_\ell e^{-v_\ell},
\]
and noticing that \(\log(q_k/p_k)+v_k=-\log Z\), the right-hand side of \eqref{eq:mixture-trial-cost} equals \(-\log Z\), proving \eqref{eq:relative-entropy-mixture}.
\end{proof}

\begin{remark}[Consequences for quantum relative entropies]
\label{rem:relative-entropy-mixture-consequences}
For faithful states, Proposition~\ref{prop:mixture-bounds} gives the logarithmic mixture bound \eqref{eq:relative-entropy-mixture} for the Umegaki, geodesic, and Belavkin--Staszewski relative entropies in \(\rho\), with \(\sigma\) fixed. The reference choices are \(\gamma=I/d\), \(\gamma=\sigma^t/\operatorname{Tr}[\sigma^t]\), and \(\gamma=\sigma\), respectively, with \(t\) held fixed in the geodesic case. Each reference therefore remains fixed while \(\rho\) is varied.
\end{remark}

At fixed \(\gamma\), joint convexity of \(V_f\) follows from that of Petz quasi-entropies \citep{petz1986quasi}, since \(A,B\) depend linearly on \(\rho,\sigma\). The logarithmic bound \eqref{eq:relative-entropy-mixture} follows from \eqref{eq:relative-entropy-minimum} and the concavity of \(X\mapsto\exp\operatorname{Tr}[\sigma\log X]\), a consequence of \citep[Theorem 3]{lieb2005stronger}. Its classical two-component form appears in \citep[Lemma 1]{ardeshiri2015mixture}, using the auxiliary-weight method of \citep{hershey2007approximating}; the geodesic specialization also follows from \citep[Proposition 4.6 and Corollary 6.3]{capel2026geodesic}. The proof above realizes the auxiliary weights \(\pvec{q}\) as probabilities for mixing the reference preparations.

\section{Retrodictive tests with a quantum channel}\label{sec:channel-extension}

To include dynamics between preparation and measurement, let the two devices prepare an input and an output system, respectively. Their reference states are connected by a quantum channel, which is used in the forward process; its Petz transpose is used in the reverse process.

Let \(\mathcal E:\mathcal B(\mathcal H_{\mathrm{in}})\to\mathcal B(\mathcal H_{\mathrm{out}})\) be a quantum channel between finite-dimensional systems, and fix faithful input states \(\rho,\gamma\). Put \(\eta=\mathcal E(\gamma)\) and restrict the output space to \(\operatorname{supp}\eta\). This loses no channel outputs: since \(\gamma>0\), every positive input \(X\) satisfies \(X\leq c\gamma\) for some \(c>0\), and hence \(\mathcal E(X)\leq c\eta\). Let \(\sigma\) be a faithful density operator on this output support, prescribed independently of \(\mathcal E(\rho)\). The input and output dimensions may differ. Define
\begin{align}
A&=\gamma^{-1/2}\rho\gamma^{-1/2},
\qquad
B=\eta^{-1/2}\sigma\eta^{-1/2},
\qquad
\mathcal R^\gamma_{\mathcal E}(X)
=\sqrt\gamma\,\mathcal E^\dagger(\eta^{-1/2}X\eta^{-1/2})\sqrt\gamma.
\label{eq:channel-relative-densities}
\end{align}
Here \(\mathcal E^\dagger\) is the trace adjoint, and \(\mathcal R^\gamma_{\mathcal E}\) is the Petz transpose of \(\mathcal E\) relative to \(\gamma\). It is a channel on the stated output support, with \(\mathcal R^\gamma_{\mathcal E}(\eta)=\gamma\).

An admissible implementation now consists of a reverse test \((\mathcal C,\pvec{a},\pvec{g})\) for \((\rho,\gamma)\) on the input and a reverse test \((\mathcal D,\pvec{b},\pvec{h})\) for \((\sigma,\eta)\) on the output. Equivalently, choose finite POVMs \(\{E_i\}\) on the input and \(\{H_j\}\) on the output support, with nonzero effects, and nonnegative weights satisfying \(A=\sum_i\lambda_iE_i\), \(B=\sum_j\mu_jH_j\). The preparation data are given by \eqref{eq:retrodictive-preparations}, with \(\gamma\) replaced by \(\eta\) on the output side.

The forward process inserts \(\mathcal E\) between preparation by \(\mathcal C\) and measurement of \(\{H_j\}\), while the reverse process inserts \(\mathcal R^\gamma_{\mathcal E}\) between preparation by \(\mathcal D\) and measurement of \(\{E_i\}\). Both retain the pair \((i,j)\). Petz transposition ensures that
\begin{align*}
J_{ij}
&=\operatorname{Tr}\!\left[
H_j\mathcal E(\sqrt\gamma E_i\sqrt\gamma)
\right]\\
&=h_j\operatorname{Tr}\!\left[
E_i\mathcal R^\gamma_{\mathcal E}(\upsilon_j)
\right],
\qquad
F_{ij}=\lambda_iJ_{ij},\quad R_{ij}=\mu_jJ_{ij}.
\end{align*}
The reference coupling \(J\) has marginals \(\pvec{g},\pvec{h}\), so its conditional laws are again related by classical Bayes' rule. We minimize over the two reverse tests at fixed \(\mathcal E,\rho,\sigma,\gamma\) and \(f\).

\begin{theorem}[Spectral optimality with a quantum channel]\label{thm:channel-optimality}
Let \(\mathcal E\) be a quantum channel between finite-dimensional systems, let \(\rho,\gamma>0\) be input density operators, let \(\eta=\mathcal E(\gamma)\), and let \(\sigma\) be a faithful density operator on \(\operatorname{supp}\eta\). Define \(A,B\) by \eqref{eq:channel-relative-densities}. For any operator-convex \(f:(0,\infty)\to\mathbb R\) with \(f(1)=0\), the minimum over the admissible channel implementations just defined is
\begin{align}
\min D_f(R\Vert F)
&=\sum_{k,l}\alpha_k f\!\left(\frac{\beta_l}{\alpha_k}\right)
\operatorname{Tr}\!\left[
Q_l\mathcal E(\sqrt\gamma P_k\sqrt\gamma)
\right],
\label{eq:channel-minimum}
\end{align}
where \(A=\sum_k\alpha_kP_k\) and \(B=\sum_l\beta_lQ_l\) are the full spectral decompositions. The spectral implementation attains the minimum simultaneously for every such \(f\), with the boundary conventions of \eqref{eq:classical-f-divergence}.
\end{theorem}

\begin{proof}
The two Jensen steps in Section~\ref{sec:spectral-optimality} still apply. The first is now traced against \(\mathcal E(\sqrt\gamma E_i\sqrt\gamma)\geq0\), and gives
\begin{align*}
&D_f(R\Vert F)\geq
\sum_l\operatorname{Tr}\!\left[
\sqrt\gamma\,\mathcal E^\dagger(Q_l)\sqrt\gamma
\sum_i\lambda_i f(\beta_l/\lambda_i)E_i
\right].
\end{align*}
Since \(\sqrt\gamma\,\mathcal E^\dagger(Q_l)\sqrt\gamma\geq0\), the second inequality \eqref{eq:second-jensen} proves \eqref{eq:channel-minimum}. Spectral choices attain equality, and zero weights are included by the same \(\varepsilon\)-limit as in Section~\ref{sec:spectral-optimality}.
\end{proof}

\begin{corollary}[The two orientations of relative entropy]
Under the assumptions of Theorem~\ref{thm:channel-optimality}, both minima are attained by the same spectral implementation and satisfy
\begin{align}
\min D(R\Vert F)
&=D_{\mathrm{BS}}(\sigma\Vert\eta)
-\operatorname{Tr}\!\left[
\mathcal R^\gamma_{\mathcal E}(\sigma)\log A
\right],\notag\\
\min D(F\Vert R)
&=D_{\mathrm{BS}}(\rho\Vert\gamma)
-\operatorname{Tr}\!\left[\mathcal E(\rho)\log B\right].
\label{eq:channel-relative-entropies}
\end{align}
\end{corollary}

\begin{proof}
Take \(f(u)=u\log u\) and \(f(u)=-\log u\), respectively, in \eqref{eq:channel-minimum}. The terms containing a single logarithm reduce using \(\sum_k\alpha_k\sqrt\gamma P_k\sqrt\gamma=\rho\) and \(\mathcal R^\gamma_{\mathcal E}(\sigma)=\sqrt\gamma\,\mathcal E^\dagger(B)\sqrt\gamma\), while \(\operatorname{Tr}[\eta B\log B]=D_{\mathrm{BS}}(\sigma\Vert\eta)\) and \(\operatorname{Tr}[\gamma A\log A]=D_{\mathrm{BS}}(\rho\Vert\gamma)\).
\end{proof}

For the identity channel, these formulas recover the preceding results without intervening dynamics. For channels with faithful Choi operator, the second line of \eqref{eq:channel-relative-entropies} agrees with the average entropy production in \citep[Eq.~(20)]{bai2026entropy}. Theorem~\ref{thm:channel-optimality} therefore characterizes that quantity as the minimum of the forward relative entropy \(D(F\Vert R)\). This identification does not extend to arbitrary channels, although the variational theorem does. Indeed, for a unitary channel the entropy production in \citep[Eq.~(23)]{bai2026entropy} vanishes for every choice of the states, whereas for the identity channel with \(\gamma=I/d\), our forward minimum is \(D(\rho\Vert\sigma)\), which is positive when \(\rho\neq\sigma\).

\section{Discussion}\label{sec:coupled-reverse-tests}

The standard, maximal, and geodesic quantum divergences thus arise from a common classical variational principle: the reference constrains the preparation ensembles and determines their retrodictive coupling. For a general reference, the minimum compares the complete forward and reverse processes and need not vanish when their prescribed quantum averages coincide.

The mixture argument suggests a concrete direction for further work. Which transformations of the states, reference, and, when present, channel can be realized by converting admissible tests into admissible tests through a common stochastic processing of the two joint laws? A conversion defined for every admissible implementation would imply, by classical data processing, that the transformed minimum cannot exceed the original one. Characterizing such conversions would therefore give conditions for monotonicity directly in terms of the permitted preparation experiments.

\section*{Acknowledgments}

I thank Mil\'an Mosonyi and Ludovico Lami for valuable comments on the manuscript. Support from MEXT Quantum Leap Flagship Program (MEXT QLEAP) Grant No.~JPMXS0120319794 and from JSPS KAKENHI Grants No.~23K03230 and~26K00621 is acknowledged. ChatGPT (OpenAI, Astra 6 and Sol 6.1) and Claude (Anthropic, Opus 5.5) were used, under the author's direction, to assist with literature searches, the examination of mathematical arguments, and the preparation and revision of the manuscript. The author takes full responsibility for the content of this work.

\bibliographystyle{retrodiction-alpha}
\bibliography{library-f-div}

\end{document}